# On the theoretical prediction of microalgae growth for parallel flow

Chunyang Ma*

*School of Advanced Manufacturing, Nanchang University, 999 Xuefu Avenue, Nanchang 330031, People's Republic of China*

*School of Mechanical and Electrical Engineering, Nanchang University, 999 Xuefu Avenue, Nanchang 330031, People's Republic of China*

**Abstract**

The established microalgae growth models are semi-empirical or considerable fitting coefficients exist currently. Therefore, the ability of the model prediction is reduced by the numerous fitting coefficients. Furthermore, the predicted results of the established models are dependent on the size of the photobioreactor (PBR), light intensity, flow and concentration field. The growth mechanism of microalgae has not clearly understood in PBR cultivation. It is difficult to predict the microalgae growth by theoretical methods, owing to the aforementioned factors. We developed an exploratory bridging microalgae growth model to predict the microalgae growth rate in PBRs by using the nondimensional method which is effectively in fluid dynamics and heat transfer. The analytical solution of the growth rate was obtained for the parallel flow. The nondimensional growth rate expressed as function of Reynolds number and Schmidt number, which can be used for arbitrary parallel flow due to the solution was expressed as nondimensional quantities. The theoretically predicted growth rate is compared with the experimentally measured microalgae growth rate on the order of magnitude. The nondimensional method successfully applied to the microalgae growth problem for the first time. The general nondimensional solution can unify the numerous experimental data for different laboratory conditions, and give a direction for the disorder of the microalgae growth problem. The nondimensional solution may be useful to explain the growth mechanism of microalgae and design large-scale PBRs for microalgae biofuel production. The significance of the work is to give a theoretical foundation and methodology of biological theory of microalgae growth.

*Keywords:* Microalgae; Nondimensional method; Parallel flow; Photobioreactors; Biofuel

---

*Corresponding author. E-mail: cyma@ncu.edu.cn, cymaphy@163.com.

## 1. Introduction

The growth of microalgae is affected by many factors such as light intensity, nutrients, and temperature. Therefore, the growth of microalgae is an extremely complex interdisciplinary problem [1, 2]. This article is the first attempt to use the dimensionless method to study the growth of microalgae for the parallel flow over a plate. The first attempt to use the dimensionless method to unify the chaotic state of microalgae growth research, just as the Reynolds number proposed in the fluid mechanics of the year. The only assumption is that the microalgae concentration satisfies the mass concentration conservation principle.

Enormous microalgae growth models have been established to solve the problem of the prediction of microalgae growth [3-5]. The existing microalgae growth models as function of a single substrate [6], such as carbon C, nitrogen N, phosphorus P, or a single light intensity [7], based on this, growth models for a function of multiple factors also proposed [8, 9]. Overall, in the growth models of microalgae, the effect of fluid flow and mass concentration diffusion process are not taken into consideration. The growth of microalgae is an extremely complex problem, which affected by many factors, such as fluid flow, nutrient concentration, light intensity/spectral, and PBR size etc.

Moreover, the growth models are semi-empirical or considerable fitting coefficients exist in the theoretical model. The aforementioned reasons cause the models to lose its generality and the prediction ability of the model is reduced. It is necessary to give a general solution to predict the growth of microalgae by following steps: (1) All the factors affecting the growth of microalgae should be considered; (2) derivation of the governing equations affect the growth of microalgae and dimensionless treatment of the equation to obtain the nondimensional solution. The greatest advantage of the nondimensional solution is its generality, i.e., the solution is independent of the PBR size, fluid type or state, nutrient concentration field and microalgae strains etc. And it is connected the fluid field, nutrient concentration field, and light field which have important effect on the growth of microalgae. In this paper, we will give a nondimensional solution of the problem of microalgae growth prediction for the parallel flow in section 3.

This study specifically aims to derivate the relation between growth rate and the influential factors, including the light, nutrients, fluid flow etc. The analytical solution of the microalgae growth rate was firstly obtained for the parallel flow. The microalgae growth rate was expressed as function of Reynolds number and Schmidt number, which can be used for arbitrary parallel flow due to the solution was expressed as dimensionless quantity. The significance of the work is to give a theoretical foundation and methodology of biological theory of microalgae growth. The results may be useful to explain the growth mechanism of microalgae and design large-scale PBRs for the microalgae cultivation.

## 2. Problem statement and methods

### *2.1 Light transfer through microalgae suspensions*

Light transfer within absorbing, scattering and non-emitting microalgal suspension in PBRs is governed by the radiative transfer equation (RTE) expressed as [10]

$$\mathbf{s}\cdot\nabla I(\mathbf{r},\mathbf{s})+\beta I(\mathbf{r},\mathbf{s})=\frac{\kappa_s}{4\pi}\int_{\Omega'=4\pi} I(\mathbf{r},\mathbf{s})\Phi(\mathbf{s}',\mathbf{s})\mathrm{d}\Omega' \tag{1}$$

where $I$ is the radiative intensity in direction $\mathbf{s}$ at location $\mathbf{r}$, $\mathbf{s}$ is the direction vector, $\kappa_s$ is the scattering coefficient, $\beta = \kappa_a + \kappa_s$ is the extinction coefficient, $\kappa_a$ is the absorption coefficient, $\Phi(\mathbf{s}',\mathbf{s})$ is the scattering phase function and $\Omega'$ is the solid angle. The scattering phase function $\Phi(\mathbf{s}',\mathbf{s})$ represents the probability that the radiation transfer in the solid angle $\mathrm{d}\Omega'$ around the direction $\mathbf{s}'$ scattered into the solid angle $\mathrm{d}\Omega$ around the direction $\mathbf{s}$ and is normalized such that

$$\frac{1}{4\pi}\int_{4\pi}\Phi(\mathbf{s}',\mathbf{s})\mathrm{d}\Omega = 1 \tag{2}$$

The asymmetry factor $g$ defined as the mean cosine of the scattering phase function expressed as [10]

$$g = \frac{1}{4\pi}\int_{4\pi}\Phi(\mathbf{s}',\mathbf{s})\cos\theta\mathrm{d}\Omega \tag{3}$$

where $\theta$ is the angle between the $\mathbf{s}$ and $\mathbf{s}'$. It was observed the microalgal suspensions featured strongly forward scattering, and its $g$ value around 0.97. Bidigare et al. used a predictive method to determine the spectral absorption coefficient $\kappa_{a,\lambda}$ by expressing it as [11]

$$\kappa_{a,\lambda} = \sum_{i=1}^{n} Ea_{\lambda,i}c_i \tag{4}$$

where $Ea_{\lambda,i}$ ($m^2 \cdot kg^{-1}$) is the *in vivo* pigment specific spectral absorption cross-section of pigment $i$, and $c_i$ ($kg \cdot m^{-3}$) is the mass concentration of the cell's pigment $i$. The specific absorption cross-section of Chl a, b and c, and $\beta$-carotene have been reported by Bidigare et al. in the spectral region 400 to 750 nm [12]. Besides, the temporal evolution spectral absorption and scattering cross-sections were experimentally measured in Refs. [13-15].

### *2.2 Nutrient supply of microalgae*

Photosynthesis is a universal process through which plants, algae, and some photosynthetic bacteria can store solar energy, electron source and carbon source (mainly $CO_2$) in the form of sugars [16]. The microalgae can produce oxygen, carbohydrates, proteins, and lipids within the cells and some species may produce $H_2$ through the photosynthesis process. The main nutrient elements used in the microalgae culture include C, N, P, and some minor elements such as potassium, magnesium, calcium, and sulfur etc. [1]. The requirements of microalgae culture can be determined from the elemental composition of the biomass. The nutrient supply should be larger than the nutrient requirements, i.e., under nutrient-sufficient conditions to maximize the productivity of the microalgae cultures.

The mass transfer must occurred which is due to the difference in the concentration of the culture, such as C, N, and P, caused by biochemical process. The transfer rate for mass diffusion is known as Fick's law [17]

$$\mathbf{j}_i = -D_{ij}\nabla\rho_i \tag{5}$$

where the subscript $i$ represent species $i$ in a mixture of $i$ and $j$, here, the $j$ maybe monocomponent or multicomponent. The quantity $\mathbf{j}_i$ is the diffusive mass flux of species $i$. And $D_{ij}$ is defined as the mass diffusion coefficient, $\rho_i$ stands for the mass concentration of species $i$ of the culture. From the species conservation, the convection diffusion equation can be expressed as [17]

$$\frac{\partial\rho_i}{\partial t} + u\frac{\partial\rho_i}{\partial x} + v\frac{\partial\rho_i}{\partial y} = D_{ij}\left(\frac{\partial^2\rho_i}{\partial x^2} + \frac{\partial^2\rho_i}{\partial y^2}\right) + \dot{\Phi}_i \tag{6}$$

where the term $\dot{\Phi}_i$ is the generation rate of the mass of species $i$ per unit volume. The concentration field can be obtained by solving the convection diffusion equation, then, to calculate the growth rate of microalgae cells by using the element balance method.

### *2.3 Fluid flow*

The microalgae culture are usually in a flow state to prevent the sedimentation of microalgae cells. In addition, in order to ensure light intensity distribution, minimize the gradients of nutrients, and maintain uniform pH, the mixing method is used in PBRs. Culture mixing may be achieved by methods, such as stirring, air bubbling, or liquid circulation, which are known to considerably enhance microalgae productivity [1]. The dynamics of the incompressible flow is governed by the Navier-Stokes equation [18]

$$\frac{\partial \vec{V}}{\partial t} + \left(\vec{V} \cdot \nabla\right)\vec{V} = -\frac{1}{\rho}\nabla p + \nu\nabla^2\vec{V} + \vec{F} \tag{7}$$

which was first proposed by Navier and further completed by Stokes, the first term about time will be disappeared for steady state flow. The vector $\vec{V}$ represents the fluid velocity, $\rho$ is the density of the fluid, $p$ is the pressure, $\nu$ is the kinematic viscosity of the fluid, and $\vec{F}$ is the mass force. The N-S equation is a second order partial differential equation, which its solution is still an open problem to be solved. Fortunately, we do not need to solve the N-S equation, the boundary layer form of the N-S equation will be used in the next section. And our aim is to find the relation between the microalgae growth and flow state of the culture.

## 3. Laminar flow over a plate

### *3.1 Assumptions and simplified conditions*

The major growth parameters may be obtained by solving the boundary layer equations. Assuming steady, incompressible, laminar flow with constant fluid properties and negligible viscous dissipation, the boundary layer equations can be written as [17]

$$\frac{\partial u}{\partial x} + \frac{\partial v}{\partial y} = 0 \tag{8}$$

$$u\frac{\partial u}{\partial x} + v\frac{\partial u}{\partial y} = \nu\frac{\partial^2 u}{\partial y^2} \tag{9}$$

$$u\frac{\partial \rho_i}{\partial x} + v\frac{\partial \rho_i}{\partial y} = D_{ij}\frac{\partial^2 \rho_i}{\partial y^2} \tag{10}$$

Assuming the microalgae cells are growth on the surface of the thin layer, the microalgae cells enter the microalgae culture through diffusion process. And the microalgae concentration, $N$, satisfy the mass concentration conservation principle is assumed. The governing equation of microalgae concentration $N$ can be similarly written as for the boundary layer

$$u\frac{\partial N}{\partial x}+v\frac{\partial N}{\partial y}=D_N\frac{\partial^2 N}{\partial y^2} \tag{11}$$

which $D_N$ represent the diffusion coefficient of microalgae cells into the mainstream culture. Solution of these equations is based on the fact that for constant properties in the velocity boundary layer are independent of species concentration. Therefore, we should begin by solving the continuity and velocity boundary layer equations. Once the velocity field has been obtained, solutions to Eq.(10) and Eq.(11), which depend on $u$ and $v$, may be obtained.

### *3.2 Analytical solution of parallel flow*

The solution follows the method defining a stream function $\psi\left(x,y\right)$, and the new dependent and independent variables, $f\left(\eta\right)$ and $\eta$, respectively, which are defined as [17]

$$u\equiv\frac{\partial\psi}{\partial y}\quad\text{and}\quad v\equiv-\frac{\partial\psi}{\partial x} \tag{12}$$

$$f\left(\eta\right)\equiv\frac{\psi}{u_\infty\sqrt{\nu x/u_\infty}} \tag{13}$$

$$\eta\equiv y\sqrt{u_\infty/\nu x} \tag{14}$$

Using these new variables, we can reduce the partial differential equations to ordinary differential equations. After some mathematical calculations, the Eq.(9) can be rewritten as by the new variables [17]

$$2\frac{d^3 f}{d\eta^3}+f\frac{d^2 f}{d\eta^2}=0 \tag{15}$$

The Eq.(15) with appropriate boundary condition can be solved by numerical integration, the important result is as follows [17]

$$\delta=\frac{5}{\sqrt{u_\infty/\nu x}}=\frac{5x}{\sqrt{Re_x}} \tag{16}$$

which $\delta$ represents the velocity boundary layer thickness. For the integrity of the results, the local friction coefficient is also given as following [17]

$$C_{f,x} = \frac{\tau_{s,x}}{\rho u_\infty^2 / 2} = 0.664 Re_x^{-1/2} \tag{17}$$

where the $\tau_{s,x}$ is the shear stress on the plate. From the solution of the velocity boundary layer, the species continuity equation can now be solved. To solve the Eq.(10) we introduce the nondimensional species density [17]

$$\rho_i^* \equiv \frac{\rho_i - \rho_{i,s}}{\rho_{i,\infty} - \rho_{i,s}} \tag{18}$$

and the boundary condition of the species density for a fixed surface

$$\rho_i^*(0) = 0 \quad \text{and} \quad \rho_i^*(\infty) = 1 \tag{19}$$

Making the necessary substitutions, the Eq.(10) can be reduced to [17]

$$\frac{d^2 \rho_i^*}{d\eta^2} + \frac{Sc}{2} f \frac{d\rho_i^*}{d\eta} = 0 \tag{20}$$

For the Schmidt number $Sc \geq 0.6$, the results of the surface species concentration gradient can be correlated by the following relation [17]

$$\left. \frac{d\rho_i^*}{d\eta} \right|_{\eta=0} = 0.332 Sc^{1/3} \tag{21}$$

The local mass transfer convection coefficient is defined as

$$h_m = \frac{n_{i,s}}{\rho_{i,s} - \rho_{i,\infty}} \tag{22}$$

which $n_{i,s}$ is the mass transfer rate of species $i$, using the Fick's law $n_{i,s}$ can be expressed as

$$n_{i,s} = -D_{ij} \left. \frac{\partial \rho_i}{\partial y} \right|_{y=0} \tag{23}$$

and combining this with Eq.(22), we obtain

$$h_m = \frac{-D_{ij} \, \partial \rho_i / \partial y \big|_{y=0}}{\rho_{i,s} - \rho_{i,\infty}} \tag{24}$$

Using the definition of $\rho_i^*$, it follows that

$$h_m = -\frac{\rho_{i,\infty} - \rho_{i,s}}{\rho_{i,s} - \rho_{i,\infty}} D_{ij} \left. \frac{\partial \rho_i^*}{\partial y} \right|_{y=0} \tag{25}$$

and with using the variable $\eta$, can be rewritten as

$$h_m = D_{ij} \left( \frac{u_\infty}{\nu x} \right)^{1/2} \left. \frac{d\rho_i^*}{d\eta} \right|_{\eta=0} \tag{26}$$

It follows that the local Sherwood number is of the form [17]

$$Sh_x = \frac{h_{m,x} x}{D_{ij}} = 0.332 Re_x^{\ 1/2} Sc^{1/3} \tag{27}$$

for the Schmidt number $Sc \geq 0.6$. From the solution to Eq.(20), the ratio of the velocity to concentration boundary layer thickness is [17]

$$\frac{\delta}{\delta_c} \approx Sc^{1/3} \tag{28}$$

where $\delta_c$ is the boundary layer thickness of the concentration, and $\delta$ is given by Eq.(16).

The boundary layer equation of microalgae concentration, Eq.(11) is of the same form as the species concentration boundary layer equation Eq.(10), with $D_N$ replacing $D_{ij}$. Introducing a nondimensional microalgae concentration $N^* \equiv N - N_s / N_\infty - N_s$ , and the appropriate boundary conditions are $N^*(0) = 0$ and $N^*(\infty) = 1$. We also see that the microalgae concentration boundary conditions are of the same form as the species boundary conditions. The mass and microalgae cells transfer analogy can be applied since the differential equation and boundary conditions for the microalgae concentration are of the same form as for species concentration. Hence, with reference to the solution of the species equation

$$Sh_x^N = \frac{h_{N,x} x}{D_N} = 0.332 Re_x^{\ 1/2} Sc_N^{1/3} \tag{29}$$

By analogy to Eq.(28), it also follows that the ratio of boundary layer thickness is

$$\frac{\delta}{\delta_N} \approx Sc_N^{1/3} \tag{30}$$

where $\delta_N$ is the boundary layer thickness of the microalgae concentration. The foregoing results can be used to compute average laminar boundary layer parameters for $x \in [0, L]$, where $L$ is the distance from the leading edge at which transition begins. The three boundary layers experience nearly identical growth for values of $Sc$ and $Sc_N$ close to unity, from Eq.(28) and Eq.(30).

From the foregoing local results, the average boundary layer parameters can be obtained by integrating the local formulas. With the average mass transfer coefficient defined as [17]

$$\overline{h_{m,x}} = \frac{1}{x}\int_0^x h_{m,x} dx \tag{31}$$

for an dimensional parallel flow. Integrating and substituting from Eq.(27) and Eq.(29), respectively, it follows that $\overline{h_{m,x}} = 2h_{m,x}$. Thus, we obtain [17]

$$\overline{Sh_x} = \frac{\overline{h_{m,x}}x}{D_{ij}} = 0.664Re_x^{1/2}Sc^{1/3} \tag{32}$$

Similarly, employing the analogy of mass and microalgae cells transfer, it follows that

$$\overline{Sh_x^N} = \frac{\overline{h_{N,x}}x}{D_N} = 0.664Re_x^{1/2}Sc_N^{1/3} \tag{33}$$

If the flow is laminar over the entire surface, the subscript $x$ can be replaced by $L$.

***3.3 Radiation energy density within the culture***

In the previous, we have already given the energy balance for thermal radiation for an infinitesimal pencil of rays, the radiative energy equation can be obtained from the integration of the radiative transfer equation over all solid angles, which as follows [10]

$$\nabla \cdot \overrightarrow{q_\lambda} - 4\pi\kappa_{a,\lambda}I_{b,\lambda} + \kappa_{a,\lambda}G_\lambda = 0 \tag{34}$$

where the $G_\lambda$ is the spectral radiation fluence rate, and $\overrightarrow{q_\lambda}$ is the spectral radiative heat flux. The self-emission term can be neglected due to the emission wavelength not in the visible region, and when applied to the flow problem the convection term should be added, which can be written as follows

$$\nabla \cdot \overrightarrow{q_\lambda} + \frac{u}{c}\frac{\partial G_\lambda}{\partial x} + \frac{v}{c}\frac{\partial G_\lambda}{\partial y} + \kappa_{a,\lambda}G_\lambda = 0 \tag{35}$$

we cannot obtain the analytical solutions of the partial differential Eq.(35). However, some simplifications can be made, the velocity of the fluid is much smaller than the light velocity $c$, the convection term can be neglected in Eq.(35), the radiative heat flux term can be simplified as $dG_\lambda / dy$ due to the scattering phase function of microalgae exhibited strongly forward scattering feature for radiation on direction of *y*, it follows that

$$\frac{dG_\lambda}{dy} + \kappa_{a,\lambda} G_\lambda = 0 \tag{36}$$

The analytical solution can be obtained as

$$G_\lambda = G_\lambda(0)\exp\left(-\kappa_{a,\lambda} y\right) = G_\lambda(0)\exp\left(-\tau y^*\right) \tag{37}$$

where the dimensionless $y^* = y/s$, and $\tau_\lambda = \kappa_{a,\lambda} s$ is the optical thickness, $s$ is the optical path length. Similarly, we introduce the concept of radiation boundary layer, which is the region that the radiative fluence rate gradients exist, and its thickness is defined as the value for which $\left(G_\lambda - G_{\lambda,s}\right)/\left(G_{\lambda,\infty} - G_{\lambda,s}\right) = 0.99$. For the radiation on direction of *x*, the boundary layer equation of radiation flux can be written as

$$\left(1 + \frac{u}{c}\right)\frac{dG_\lambda}{dx} + \kappa_{a,\lambda} G_\lambda = 0 \tag{38}$$

If the velocity $u = u\left(y\right)$, which may be a reasonable assumption for the parallel flow, the solution can be expressed as

$$G_\lambda\left(x\right) = G_\lambda\left(0\right)\exp\left[-\kappa_{a,\lambda} x \middle/ \left(1 + \frac{u}{c}\right)\right] \tag{39}$$

The same direction of velocity $u$ will be enhanced the light intensity for a fixed point, whereas, the reverse direction of $u$ will recede the light intensity. The effect of flow velocity enhance the light intensity may be very small, however, if this effect makes rational use of us, it will effectively improve the utilization rate of light energy.

### *3.4 Growth rate of microalgae*

According to the process of microalgae cells diffusion to the flow from the surface and the microalgae cells concentration increase rate, $N_{M,s}$, in the parallel flow, it follows that

$$N_{M,s}=\int_S n_{M,s}dS=\frac{d}{dt}\int_V NdV=\int_V \frac{dN}{dt}dV=\bar{\mu}\int_{V|V_c} NdV \tag{40}$$

where the $\bar{\mu}$ is the average growth rate of microalgae culture, the $V_c$ is the volume of constant concentration of microalgae, and $V|V_c$ represents the total volume subtract the volume $V_c$. From the definition, the microalgae cells transfer rate can also be expressed as

$$N_{M,s}=\left(N_s-N_\infty\right)\int_S h_N dS \tag{41}$$

where the subscript $M$ emphasize the microalgae. Combining the Eq. (40) and Eq.(41), it follows that

$$\bar{\mu}=\frac{\left(N_s-N_\infty\right)}{\int_{V_{BL}} NdV}\int_S h_N dS \tag{42}$$

Using the relation of local and average mass transfer coefficient, the Eq.(42) can be rewritten as

$$\bar{\mu}=\frac{\overline{h_N}\left(N_s-N_\infty\right)S}{\int_{V_{BL}} NdV} \tag{43}$$

For one dimensional parallel flow, the integral can be approximated by the following

$$\bar{\mu}=\frac{\overline{h_N}\left(N_s-N_\infty\right)L}{N_\infty \delta_N^L L} \tag{44}$$

which assumed that the concentration of boundary layer is slightly difference with the concentration of mainstream. By multiplying variable $x$ and dividing $D_N$ on both sides of Eq.(44), and combine with Eq.(29), it follows that

$$\frac{\bar{\mu}x\delta_N^L}{D_N}=\overline{Sh_x^N}\frac{\left(N_s-N_\infty\right)}{N_\infty} \tag{45}$$

Substituting the expression of $\delta_N^L=5L\mathrm{Re}_L^{-1/2}Sc_N^{-1/3}$ into Eq.(45), and defining the dimensionless number $\overline{Mg_x}=\bar{\mu}xL/D_N$ which may be temporally called 'Microalgae growth number', we obtain

$$\overline{Mg_x}=\frac{1}{5}\frac{\left(N_s-N_\infty\right)}{N_\infty}\overline{Sh_x^N}\,\mathrm{Re}_L^{1/2}\,Sc_N^{1/3} \tag{46}$$

Combining the Eq. (33) we obtain,

$$\overline{Mg_x} = \frac{0.664}{5}\frac{\left(N_s - N_\infty\right)}{N_\infty} {Re_x}^{1/2} \mathrm{Re}_L^{1/2} Sc_N^{2/3} \tag{47}$$

The Microalgae growth number is linearly proportional to the Reynolds number from Eq.(47).

The growth rate can also be expressed from the nutrients substrate analysis, the consumption of nutrient elements is transformed into microalgae biomass through photosynthesis. Therefore, the nutrients supply of the culture must be larger than the nutrients assimilation of microalgae, which follows that [2]

$$\overline{n_{i,s}} A \geq \mu NV / Y_i \tag{48}$$

where $Y_i$ is the growth yield [2], and subscript $i$ represent the nutrient element, such as C, N, P. Combining with Eq.(22) and Eq.(32), we can obtain

$$\mu \leq 0.664 \frac{D_{ij}}{x} \frac{AY_i}{NV} \left(\rho_{i,s} - \rho_{i,\infty}\right) Re_x^{1/2} Sc^{1/3} \tag{49}$$

The average growth rate can be determined from Eq.(49) by an integration over the plate

$$\bar{\mu} \leq 1.328 D_{ij} \frac{AY_i}{NVL} \left(\rho_{i,s} - \rho_{i,\infty}\right) Re_L^{1/2} Sc^{1/3} \tag{50}$$

From the start point of light supply, the light energy supply must be larger than the needed energy in photosynthesis, which can be expressed as [2]

$$GA \geq \mu NV / Y_l \tag{51}$$

where $Y_l$ is the growth yield for light energy [2]. Combining with Eq.(37), is follows that

$$\mu \leq \frac{AY_l}{NV} \int_{\mathrm{PAR}} G_\lambda\left(0\right) \exp\left(-\tau y^*\right) d\lambda \tag{52}$$

where the subscript PAR in the integral stands for the photosynthetically active radiation (PAR), which over the spectral region from 400 to 700 nm. The average growth rate can be obtained as following

$$\bar{\mu} \leq \frac{AY_l}{NV} \int_{\mathrm{PAR}} \frac{G_\lambda\left(0\right)}{\tau_\lambda} \left[1 - \exp\left(-\tau_\lambda\right)\right] d\lambda \tag{53}$$

The Eq.(49) and Eq.(52) give the upper limit of the growth rate of microalgae under different perspectives. If we make the two equal, it follows that

$$0.664\frac{D_{ij}}{x}Y_i\left(\rho_{i,s}-\rho_{i,\infty}\right)Re_x^{1/2}Sc^{1/3}=Y_l\int_{\mathrm{PAR}}G_\lambda\left(0\right)\exp\left(-\tau y^*\right)d\lambda \quad (54)$$

The equation gives the connection between fluid flow, mass transfer, and light irradiance. The maximum growth rate of microalgae cultivation may be achieved according to this relation, which points out the quantitative relationship in microalgae cultivation for PBR design. It is indicated that large amount of light irradiance should be accompany with relatively large velocity and mass transfer intensity in order to facilitate the growth of microalgae.

### *3.5 Turbulent flow over an plate*

The turbulent flows will be occurred with Reynolds number exceeds the critical Reynolds number, $Re_{x,c}$. By employing the Chilton-Colburn analogy, the local Sherwood number for turbulent flow is [17]

$$Sh_x=St^*Re_xSc=0.0296{Re_x}^{4/5}Sc^{1/3} \quad (55)$$

with the condition $0.6 < Sc < 3000$. And the local Sherwood number for microalgae is

$$Sh_x^N=St_N^*Re_xSc_N=0.0296{Re_x}^{4/5}Sc_N^{1/3} \quad (56)$$

We can see that the local Sherwood number grow more rapidly than laminar boundary layer caused by enhanced mixing of turbulent boundary layer. The average coefficients can be determined for mixed boundary layer conditions with the definition as following [17]

$$\bar{h}_L=\frac{1}{L}\left(\int_0^{x_c}h_l\mathrm{d}x+\int_{x_c}^{L}h_t\mathrm{d}x\right) \quad (57)$$

where it is assumed that transition occurs abruptly at $x=x_c$. The average Sherwood number can be expressed as [17]

$$\overline{Sh_L}=\left(0.037{Re_L}^{4/5}-C\right)Sc^{1/3} \quad (58)$$

and the average Sherwood number for microalgae is

$$\overline{Sh_L^N}=\left(0.037{Re_L}^{4/5}-C\right)Sc_N^{1/3} \quad (59)$$

where the $C$ = 871 for a transition Reynolds number of $Re_{x,c} = 5\times10^5$ [17]. The average Microalgae growth number for mixed boundary layer is

$$\overline{Mg_L} = \frac{1}{5}\frac{\left(N_s - N_\infty\right)}{N_\infty}\overline{Sh_L^N}\,\mathrm{Re}_L^{1/2}\,Sc_N^{1/3} \tag{60}$$

Substituting Eq.(59) into Eq.(60), we can obtain

$$\overline{Mg_L} = \frac{1}{5}\frac{\left(N_s - N_\infty\right)}{N_\infty}\left(0.037Re_L^{\;4/5} - C\right)\mathrm{Re}_L^{1/2}\,Sc_N^{2/3} \tag{61}$$

The growth number of mixed boundary layer increases faster than the growth number of laminar boundary layer. This implies that the growth of microalgae will benefit from the enhanced mixing.

The average growth rate based on the nutrient elements balance for mixed boundary layer is

$$\bar{\mu} \le \frac{AY_iD_{ij}}{NVL}\left(\rho_{i,s} - \rho_{i,\infty}\right)\left(0.037Re_L^{\;4/5} - C\right)Sc^{1/3} \tag{62}$$

This gives the upper limit of the growth rate at the mixed boundary layer condition. It is also revealed the quantitative relation for growth rate with the flow velocity and mass transfer. Combining with the Eq.(53), the light, flow, and mass transfer formula can be obtained

$$\frac{Y_iD_{ij}}{L}\left(\rho_{i,s} - \rho_{i,\infty}\right)\left(0.037Re_L^{\;4/5} - C\right)Sc^{1/3} = Y_l\int_{\mathrm{PAR}}\frac{G_\lambda\left(0\right)}{\tau_\lambda}\left[1 - \exp\left(-\tau_\lambda\right)\right]d\lambda \tag{63}$$

This relation is valid for mixed boundary layer condition. It is indicated that the more light irradiance the more nutrient supply to promote the growth of microalgae. It is also indicated that different light spectrum will need different nutrient supply. The equation gives the basic quantitative expression for the suitable growth supply of microalgae culture.

## 4. Results and discussions

### *4.1 Prediction of the solution for parallel flow*

In the previous section, we have derived the expression of growth rate of microalgae for parallel flow.

Here, we will further analyze the relation between the Growth number and the Reynolds number, and the growth rate of microalgae varies with the Reynolds number. The calculated results are presented in Fig. 1 and Fig. 2, respectively, for laminar and mixed boundary layer condition by using Eq.(47) and Eq.(61). The critical Reynolds number chooses as $\mathrm{Re}_c = 5\times10^5$ for the external parallel flow [17]. The kinematic viscosity is $8.9\times10^{-6}$ $m^2\cdot s^{-1}$, and diffusion coefficient is $1.973\times10^{-9}$ $m^2\cdot s^{-1}$ [19]. The velocity is 0.5 $m\cdot s^{-1}$, and the length varies from 0.5 to 8.9 m for laminar flow. The ratio $R$ is defined as $N_s/N_\infty$ . As shown in Fig. 1, the varying trend abruptly changed at the point of critical Reynolds number, which is due to the transition from laminar flow to turbulent flow. The Growth number increases with the increase of Reynolds number for different ratios of $R$ .

As shown in Fig. 2, the growth rate decreases with the increase of Reynolds number for the laminar flow, and it has a tendency to increase first and then decrease trend for the mixed boundary layer condition. It is indicated that the low speed flow will be more suitable to the growth of microalgae (Note that this is valid only for laminar flow condition, and the formula given in the paper is no longer applicable for small Reynolds number). The extreme point of the Reynolds number which to achieve the maximum growth rate for mixed boundary layer can be calculated by the derivation of Eq.(61) with respect to length $L$, and make it equal to zero follows that

$$Re_L = 159.77C^{5/4} \tag{64}$$

here, the $Re_L = 7.5\times10^5$ when $C$ gets the value 871. That is to say, when the Reynolds number takes $7.5\times10^5$, the growth rate reaches the maximum for the mixed boundary layer in this paper. We can easily obtain the Reynolds number that maximizes the growth rate according to Eq.(64). Therefore, according to the kinematic viscosity and characteristic length, we can determine the optimum flow velocity.

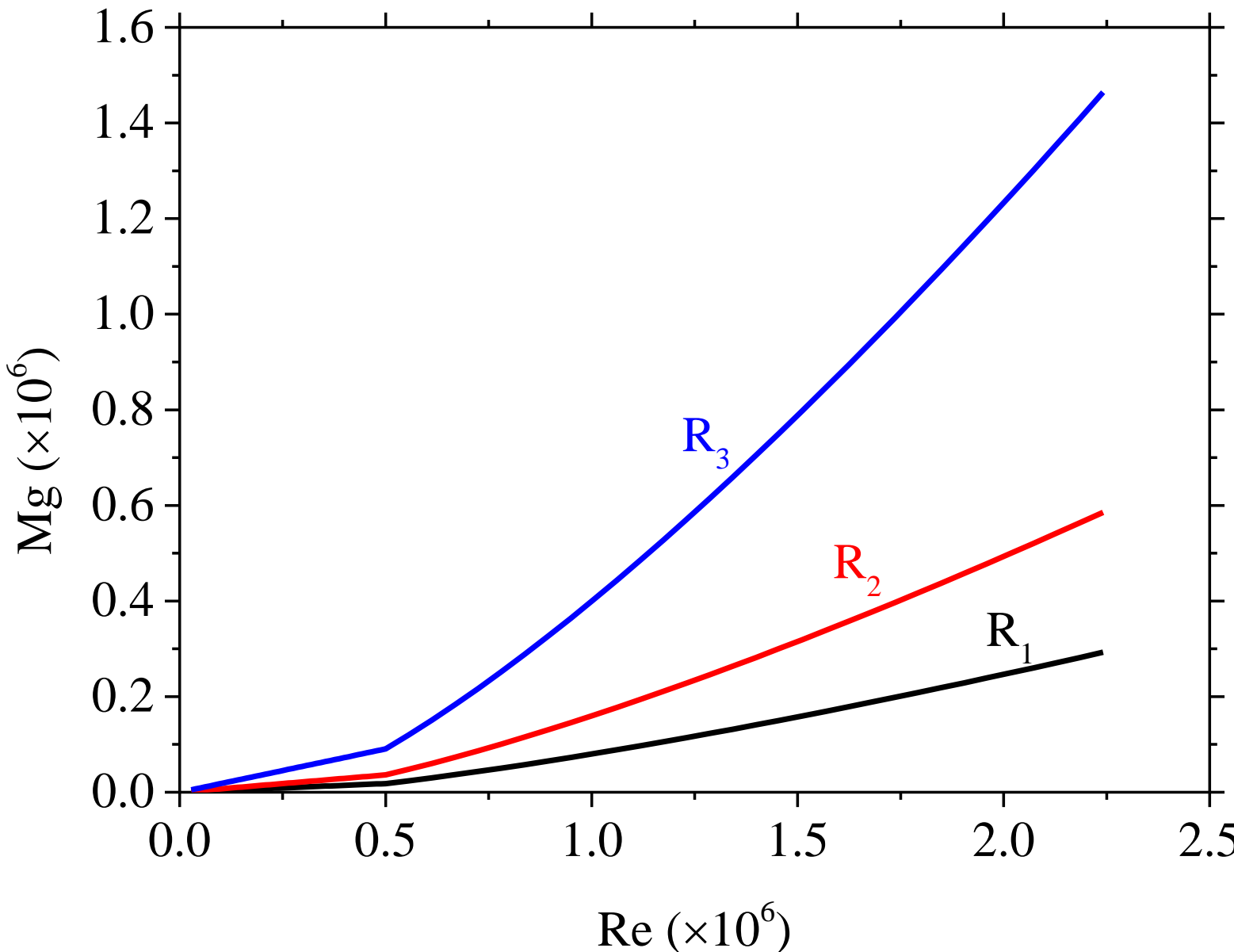

**Fig. 1.** The Growth number vs the Reynolds number, the ratio $R_1$, $R_2$, and $R_3$ equals to 1.001, 1.002, and 1.005, respectively.

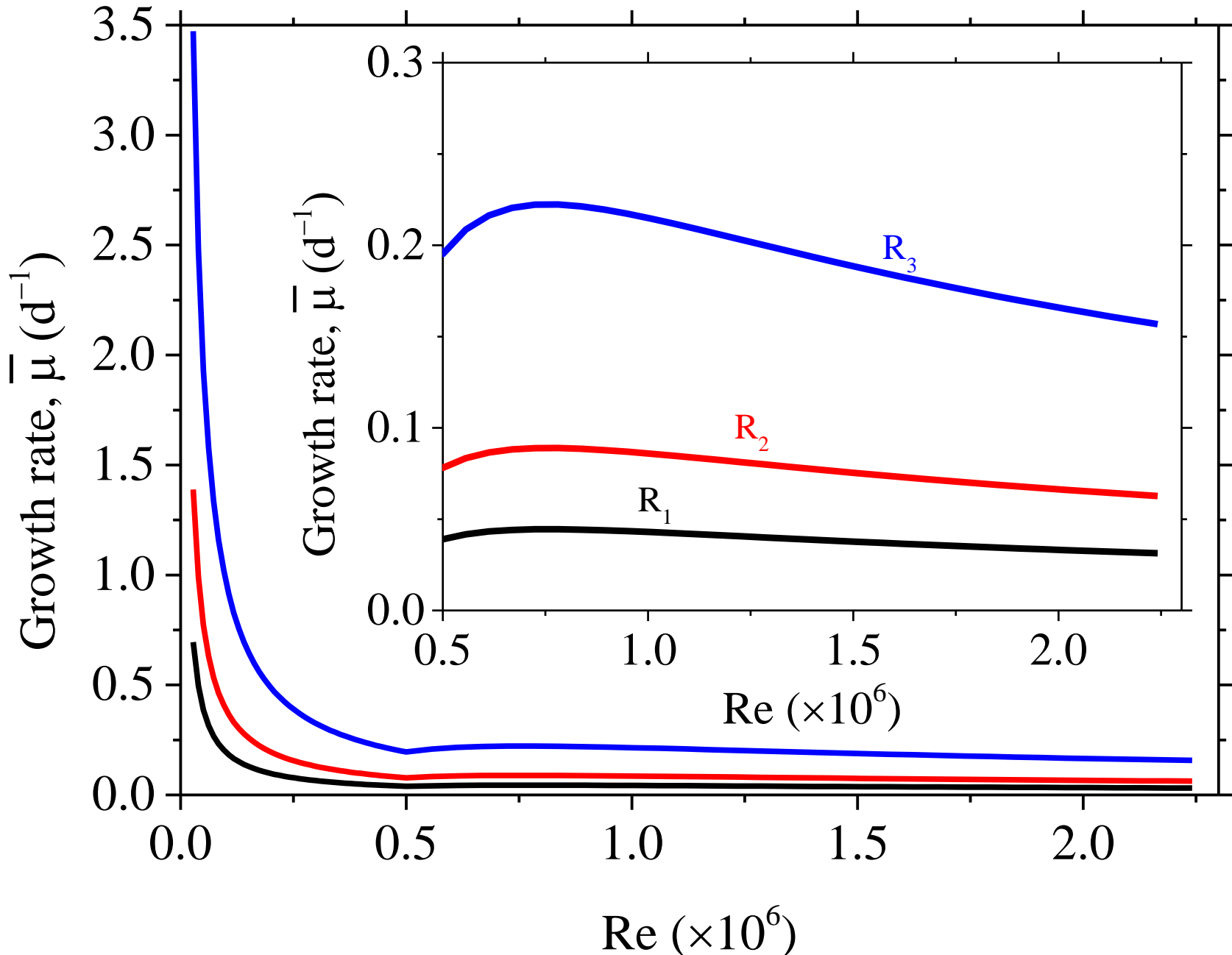

**Fig. 2.** The growth rate vs the Reynolds number, the ratio $R_1$, $R_2$, and $R_3$ equals to 1.001, 1.002, and 1.005, respectively.

The growth rate with respect to the Reynolds number for different diffusion coefficient is shown in Fig. 3. As shown, the growth rate increases with the increase of diffusion coefficient, when the Reynolds number keep invariant. It is in line with our intuition that a large diffusion coefficient corresponds to a large growth rate, which may be due to the relatively strong mass transfer capacity. Therefore, we should improve the mass transfer as much as possible in the culture of microalgae, or choose the medium with strong mass transfer capacity. Fig. 4 shows the growth rate with respect to the Reynolds number for different kinematic viscosity. As shown, the growth rate decreases with the increase of kinematic viscosity of the culture for a fixed Reynolds number. Hence, we should dilute the culture in order to reduce its viscosity in the cultivation of microalgae to improve the productivity of microalgae. It is note worthy that the change of viscosity means the change of velocity and characteristic length under the condition of the Reynolds number is constant.

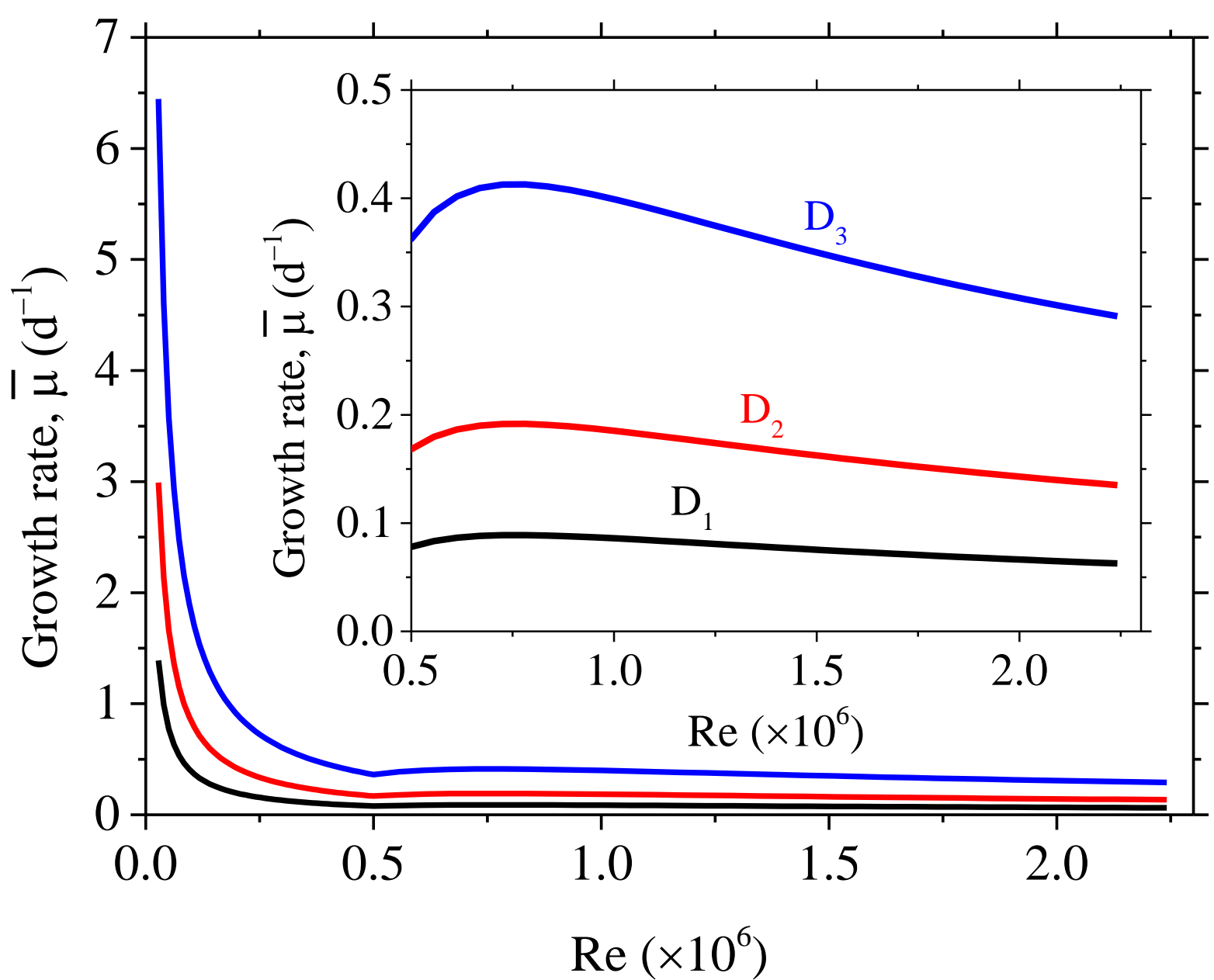

**Fig. 3**. The growth rate vs the Reynolds number for different diffusion coefficients, the ratio R equals to 1.002, the $D_1$, $D_2$, and $D_3$ equals to $1.973\times10^{-9}$ $m^2\cdot s^{-1}$, $1.973\times10^{-8}$ $m^2\cdot s^{-1}$, and $1.973\times10^{-7}$ $m^2\cdot s^{-1}$, respectively.

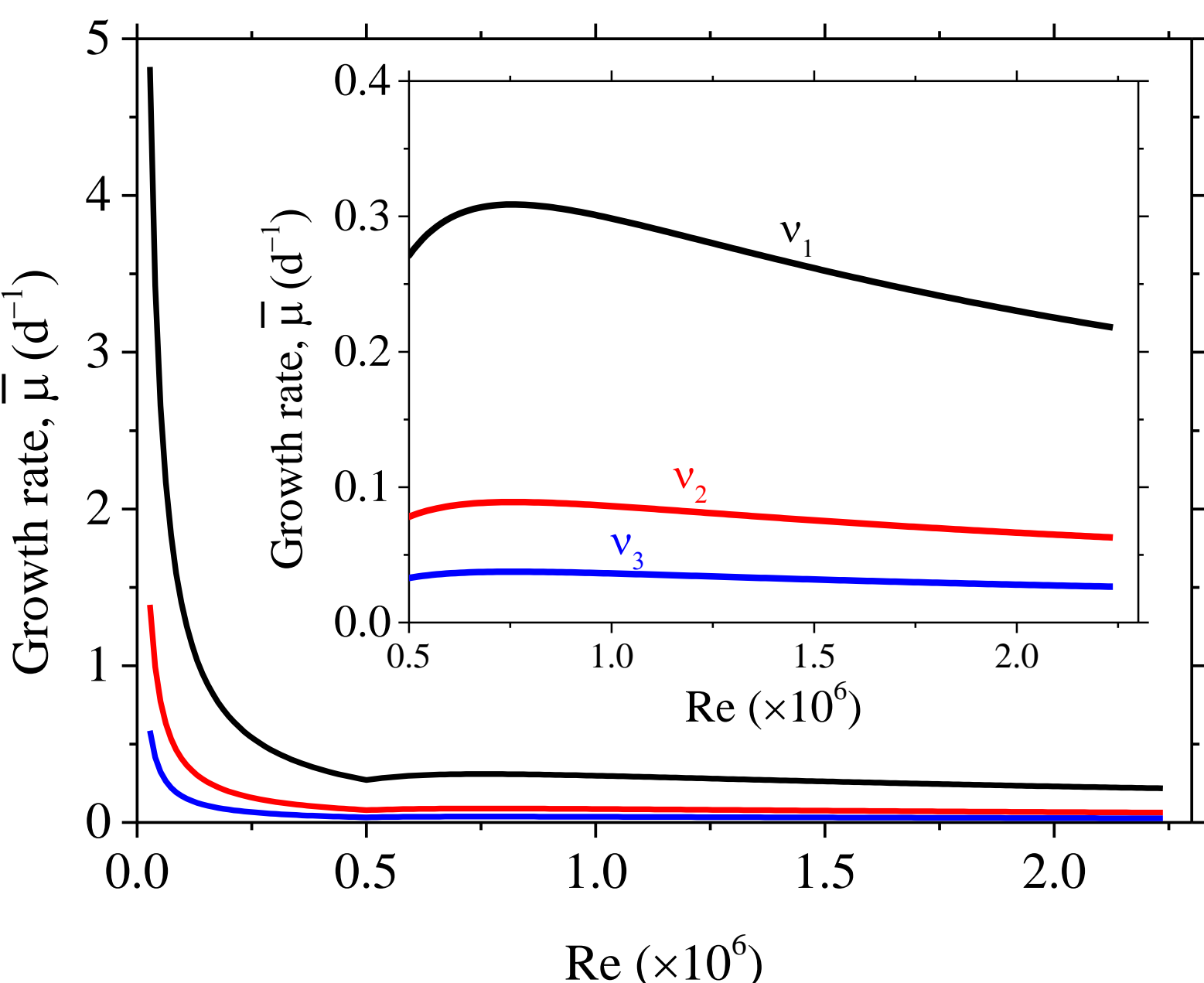

**Fig. 4.** The growth rate vs the Reynolds number for different kinematic viscosity, the ratio R equals to 1.002, the $\nu_1$, $\nu_2$, and $\nu_3$ equals to $3.5\times10^{-6}$ $m^2\cdot s^{-1}$, $8.9\times10^{-6}$ $m^2\cdot s^{-1}$, and $1.7\times10^{-5}$ $m^2\cdot s^{-1}$, respectively.

### *4.2 Experimental data from PBR operation*

The growth rate and cultivation parameters are summarized in Table 1. As shown, the range of the growth rate from 0.0465 to 1.752 $d^{-1}$ for more than 8 species of microalgae. However, the Reynolds number of three cases were obtained for species of *Scenedesmus sp.* and *Chlorella sp.*, which due to the lack of length and velocity data. The theoretically predicted growth rate (in Fig. 2) is within the range of the experimental growth rate which corresponding to the Reynolds number $8.43\times10^4$, $4.49\times10^4$, and $5.35\times10^4$, respectively. In general, the growth rate predicted by the theory is consistent with the growth rate obtained by the experimental measurement on the order of magnitude. It can be said that the semi-quantitative predictions of microalgal growth rather than accurate predictions achieved success. It is well known that precise theoretical prediction of the growth of microbial life is very difficult or even impossible. Therefore, we are trying to give a new method or a new view for explaining/understanding the growth mechanism of microalgae in this paper.

**Table 1** The growth rate of microalgae.

| Microalgae strains | Velocity, $v$ (m·s$^{-1}$) | Length, $L$ (m) | Reynolds number, $Re$ | Growth rate, $\bar{\mu}$ (d$^{-1}$) | Source |
|---|---|---|---|---|---|
| *Chlorella spp.* | 0.5 | / | / | 0.11~0.14 | Masojídek et al. [20] |
| *Scenedesmus sp.* | 0.5 | 1.5 | $8.43\times10^{4}$ | 0.14~0.24 | Doucha and Lívanský [21] |
| *C. vulgaris* and *Nannochloropsis sp.* | 0.498 | / | / | 0.14 | Torzillo et al. [22] |
| *Haematococcus pluvialis* | / | / | / | 0.6727 $^{m}$ | Zhang et al. [9] |
| *Chlorella sp.,* | 0.20 | 2.0 | $4.49\times10^{4}$ | 0.2~0.83 | Yan et al.[23] |
| *Chlorella sp.* | 0.238 | 2.0 | $5.35\times10^{4}$ | 1.75$^{a}$ | Yan et al.[24] |
| *Chrysophycea* | / | / | / | 0.0465 $^{m}$ | Malve et al. [25] |
| *Chlamydomonas acidophila* | / | / | / | 1.416~1.752 $^{m}$ | Spijkerman et al. [8] |
| *Chlorella sp.*, *Synechocystis sp. PCC 6803*, and *Tetraselmis suecica* | / | / | / | 0.984 $^{m}$ | He et al. [26] |

$^{m}$ Maximum growth rate, $^{a}$ Average growth rate.

## 5. Conclusions

This study firstly presented the analytical solution of the growth rate of microalgae for parallel flow, which considered influential factors of light, nutrients, and fluid flow. The solution gives comprehensive understanding of how the growth rate affected by the influential factors. The nondimensional growth rate of microalgae is expressed as function of Reynolds number and Schmidt number, which can be used for arbitrary parallel flow due to the solution is expressed in dimensionless form. From the solution, we can see that larger diffusion coefficient will promote the growth of microalgae, which may also explain why microalgae grow faster in solid medium. The growth rate solution can predict the experimentally measured growth rate on the order of magnitude. These results may be useful in the design and operation of PBRs for biofuel production.

## Declarations

### Ethics approval and consent to participate

No human or animal rights are applicable to this study.

### Consent for publication

Not applicable.

### Availability of data and material

The datasets during and/or analyzed during the current study available from the corresponding author on reasonable request.

All data generated or analyzed during this study are included in this published article.

### Competing interests

The author declares no competing interests.

### Funding

This work was supported by Jiangxi Provincial Natural Science Foundation (No. 20212BAB214060) and Nanchang University, which are gratefully acknowledged.

### Authors' contributions

The work presented here was completely carried out by Chunyang Ma. It includes the conception and design of the study, interpretation of the results, presentation of the article, and critical revision of the manuscript as well as final approval of the manuscript.

## Acknowledgements

Thank you for my wife, this paper wouldn't have been possible without your selfless care and unconditional love for our family.

## References

[1] Muñoz R, Gonzalez-Fernandez C. Microalgae-based biofuels and bioproducts: from feedstock cultivation to end-products. New York: Woodhead Publishing; 2017.

[2] Richmond A, Hu Q. Handbook of microalgal culture: applied phycology and biotechnology. Oxford, UK: John Wiley & Sons; 2013.

[3] Blanken W, Postma PR, de Winter L, Wijffels RH, Janssen M. Predicting microalgae growth. Algal research. 2016;14:28-38.

[4] Lee E, Jalalizadeh M, Zhang Q. Growth kinetic models for microalgae cultivation: a review. Algal Research. 2015;12:497-512.

[5] Packer A, Li Y, Andersen T, Hu Q, Kuang Y, Sommerfeld M. Growth and neutral lipid synthesis in green microalgae: a mathematical model. Bioresource technology. 2011;102:111-7.

[6] Xin L, Hong-ying H, Ke G, Ying-xue S. Effects of different nitrogen and phosphorus concentrations on the growth, nutrient uptake, and lipid accumulation of a freshwater microalga Scenedesmus sp. Bioresource Technology. 2010;101:5494-500.

[7] Bernard O, Rémond B. Validation of a simple model accounting for light and temperature effect on microalgal growth. Bioresource Technology. 2012;123:520-7.

[8] Spijkerman E, De Castro F, Gaedke U. Independent colimitation for carbon dioxide and inorganic phosphorus. PLoS One. 2011;6:e28219.

[9] Zhang X, Gong X, Chen F. Kinetic models for astaxanthin production by high cell density mixotrophic

culture of the microalga Haematococcus pluvialis. Journal of Industrial Microbiology and Biotechnology. 1999;23:691-6.

[10] Modest MF. Radiative heat transfer. New York: Academic press; 2013.

[11] Bidigare RR, Smith R, Baker K, Marra J. Oceanic primary production estimates from measurements of spectral irradiance and pigment concentrations. Global Biogeochemical Cycles. 1987;1:171-86.

[12] Bidigare RR, Ondrusek ME, Morrow JH, Kiefer DA. In-vivo absorption properties of algal pigments. International Society for Optics and Photonics. 1990;1302:290-302.

[13] Heng RL, Pilon L. Time-dependent radiation characteristics of Nannochloropsis oculata during batch culture. Journal of Quantitative Spectroscopy and Radiative Transfer. 2014;144:154-63.

[14] Ma CY, Zhao JM, Liu LH. Experimental study of the temporal scaling characteristics of growth-dependent radiative properties of Spirulina platensis. Journal of Quantitative Spectroscopy and Radiative Transfer. 2018;217:453-8.

[15] Zhao JM, Ma CY, Liu LH. Temporal scaling of the growth dependent optical properties of microalgae. Journal of Quantitative Spectroscopy and Radiative Transfer. 2018;214:61-70.

[16] Pilon L, Berberoğlu H, Kandilian R. Radiation transfer in photobiological carbon dioxide fixation and fuel production by microalgae. Journal of Quantitative Spectroscopy and Radiative Transfer. 2011;112:2639-60.

[17] Incropera FP, Lavine AS, Bergman TL, DeWitt DP. Fundamentals of heat and mass transfer. New York: Wiley; 2007.

[18] Rieutord M. Fluid dynamics: an introduction: Springer; 2014.

[19] Tropea C, Yarin AL. Springer handbook of experimental fluid mechanics. Berlin: Springer Science & Business Media; 2007.

[20] Masojídek J, Kopecký J, Giannelli L, Torzillo G. Productivity correlated to photobiochemical

performance of Chlorella mass cultures grown outdoors in thin-layer cascades. Journal of industrial microbiology & biotechnology. 2011;38:307-17.

[21] Doucha J, Lívanský K. Novel outdoor thin-layer high density microalgal culture system: Productivity and operational parameters. Algological Studies/Archiv Hydrobiologie Supplement Volumes. 1995:129-47.

[22] Torzillo G, Giannelli L, Martínez-Roldán A, Verdone N, De Filippis P. Microalgae culturing in thin-layer photobioreactors. Chemical Engineering. 2010;20:265-70.

[23] Yan C, Zhang Q, Xue S, Sun Z, Wu X, Wang Z, et al. A novel low-cost thin-film flat plate photobioreactor for microalgae cultivation. Biotechnology and Bioprocess Engineering. 2016;21:103-9.

[24] Yan C, Wang Z, Wu X, Wen S, Yu J, Cong W. Outdoor cultivation of Chlorella sp. in an improved thin-film flat-plate photobioreactor in desertification areas. Journal of Bioscience and Bioengineering. 2020;129:619-23.

[25] Malve O, Laine M, Haario H, Kirkkala T, Sarvala J. Bayesian modelling of algal mass occurrences—using adaptive MCMC methods with a lake water quality model. Environmental Modelling & Software. 2007;22:966-77.

[26] He L, Subramanian VR, Tang YJ. Experimental analysis and model-based optimization of microalgae growth in photo-bioreactors using flue gas. biomass and bioenergy. 2012;41:131-8.